\documentclass[showpacs,preprintnumbers,amsmath,amssymb]{revtex4}
\usepackage[dvips]{graphicx}

\begin{document}
\title{\Large\bf {Lower Excitation Spectrum of
the Nucleon and Delta in a Relativistic Chiral Quark Model}
}
\author{E.M. Tursunov}
\address{Institute of Nuclear Physics, Uzbekistan
 Academy of Sciences, 702132, Ulugbek, Tashkent, Uzbekistan
 e-mail: tursune@inp.uz}
\begin{abstract}
The lower excitation spectrum of
the nucleon and $\Delta$ is calculated in a relativistic chiral
quark model. Contributions of the second order self-energy and exchange diagrams
due-to pion fields to
the mass spectrum of the $SU(2)$ baryons are estimated.  A
splitting between $N(939)$ and positive parity nucleon resonance
(Roper resonance) $N^*(1440)$ is reproduced with a reasonable accuracy.
The obtained structure of one-meson exchange interaction confirms a prediction
of the large $N_c$ limit approach stating that the mass splitting between
various baryon states receive contributions from operators which
simultaneously couple spin, isospin and orbital momentum. It is shown that
one-meson exchange interaction generates a splitting
between negative parity $N^*(1/2^-)$ and $N^*(3/2^-)$ states, and also between
$\Delta^*(3/2^-)$ and $\Delta^*(1/2^-)$ states in contrast to the
non-relativistic Goldstone-Boson Exchange based quark models. This splitting is
due-to a relativistic operator which couples the lower and upper orbital
momentum of two interacting valence quarks.
\end{abstract}
\maketitle

\section{Introduction}
\par There is a growing interest to the baryon spectroscopy in the light of recent discovery of
pentaquark baryon $\Theta^+$ with a positive strangeness and mass
value $M=1540 \pm 10$ MeV and width $\Gamma < 25$ MeV by the LEPS
group at the SPRING-8 facility in Harima, Japan \cite{nac03} and
another pentaquark baryon $\Xi^{--}$ with $M=1862$ MeV and $\Gamma
< 18$ MeV by the NA49 group at CERN \cite{alt04}. The minimal
possible quark content of the $\Theta^+$ and $\Xi^{--}$ is
$u^2d^2\bar{s}$ and $s^2d^2\bar{u}$ respectively and therefore
they are called pentaquarks. The discovery of pentaquarks in
particular inspired a new overview on the light hadron
spectroscopy, the lowest excitation spectrum of the Nucleon and
$\Delta$. One new point here is that the long standing puzzle of
the Roper resonance $N^*(1440)(1/2^+)$ is suggested to be resolved
with an idea that this baryon state has a pentaquark content
$u^2d^2\bar{d}$ \cite{jaf04}. In addition to the long discussed
exotic structure of the Roper resonance \cite{li91}, the new point
requires a detailed study of the lowest excitation spectrum of
light baryons.
\par The most successful study of the spectroscopy of the lowest
excited states of the $N$ and $\Delta$ was done in the traditional
One-Gluon Exchange (OGE) Constituent Quark Model
\cite{isgur77,cap86,stan90,stan91}. Another prominent, Goldstone
Boson Exchange (GBE) Constituent Quark Model \cite{gloz96} is
based on the idea that the spontaneous breaking of the chiral
symmetry of QCD plays a crucial role for the hadron structure.
Continuing debate of the two above models over last decade
\cite{isgur99,cap00} was focused mainly on the so-called
"three-body spin orbit problem" associated with the OGE mechanism.
The authors of the GBE based model claims that the Goldstone boson
exchange, rather than the gluon exchange, is the source of the
hyperfine splitting, and that there is no any spin-orbit splitting
between for-example, negative parity $N^*(1/2^-)$ and $N^*(3/2^-)$
resonances, and also between $\Delta^*(3/2^-)$ and
$\Delta^*(1/2^-)$ resonances. Unfortunately, the current status of
the experimental data still does not allow to overcome these
problems \cite{PDG03}. Although, the new analysis of the
experimental data \cite{arndt04} confirms the splitting in above
nucleon and $\Delta$ bands.
\par On the other hand, neither of these models of the baryon
structure is relativistic and neither is QCD and it is impossible
to draw any conclusion concerning the nature of the hyperfine
interaction in baryons. It is clear now that all results of the
constituent quark models have to be reexamined in some
relativistic quark model. Up to now, the QCD based models like the
MIT bag \cite{MIT}, cloudy bag \cite{thomas81} and chiral bag
\cite{bro80}  have not been extended to the excited baryon
sector. Most detailed study of the masses of light hadrons in
cloudy and chiral bag models was done by Saito \cite{saito84},
however this work also did not include lower excitation spectrum
of the Nucleon and Delta. The inclusion of the Instanton induced
interaction mechanism in the frame of the MIT bag model
\cite{dor90} also was restricted to the octet baryons.
\par The octet and decuplet baryons also were studied successfully in a fully
relativistic Poincare covariant Diquark-Quark Model based on Bethe-Salpeter
equation \cite{al99}. Another model which respects the Poincare covariance and
which preserves all results of the original cloudy bag model \cite{thomas81} is
the Light Front Cloudy Bag Model \cite{mil02}. The model recently was
extended for the extrapolation of lattice QCD results to the physical values of
$m_{\pi}^2$ \cite{hr05}. This work have demonstrated the importance of the development
 of cavity models for hadron structure.
  On the other hand, the lattice studies still did not reproduce
the main properties of the excited $N$ and $\Delta$ sector, although the ground
states $N(939)$ and $\Delta(1232)$ are described well (see recent review \cite{lein04}).
\par The aim of the present work is to develop a relativistic chiral quark model
based on field-theoretical description of the interquark
interaction for the excited Nucleon and $\Delta$ spectroscopy. The
model was firstly suggested in Ref. \cite{oset84} and used for the
calculations of the Nucleon ground state properties in
\cite{gutsche87,gutsche89}. A modification of the model, the
so-called Perturbative Chiral Quark Model was recently extended to
the systematic study of the nucleon properties \cite{lyub01}, the
mass-spectrum of the octet-baryons \cite{inoue041} and the
pentaquark systems \cite{inoue042} . In these approaches baryons
are considered as bound states of valence quarks surrounded by a
cloud of Goldstone bosons ($\pi-, K-, \eta-$mesons )  as required
by the chiral symmetry.
\par As was found in the model independent analyses of baryons in the large $N_c$
limit \cite{schat02}, the mass splitting between various baryon states receive
important contributions from operators which simultaneously couple spin, isospin and
orbital momentum. On the other hand (see \cite{gutsche87,gutsche89}),
the relativistic structure of the one-meson
exchange forces for the ground states of the N and $\Delta$ is the
same as in the Goldstone Boson Exchange Model of Glozman et al.,
i.e. it is
\begin{equation}
V(r_{ij}) \ \hat{\vec \sigma_i}* \hat{\vec \sigma_j} \ \hat{\vec
\tau_i} *\hat{\vec \tau_j} \ ,
\end{equation}
where $\hat{\vec \sigma_i}$ and $\hat{\vec \tau_i}$ are spin and
isospin operators respectively. It is important to note that the
above structure of GBE interaction yields the correct ordering of
the radially and orbitally excited nucleon resonances, namely the
Roper resonance $N(1440)(1/2^+)$ and negative parity
$N(1520)(3/2^-)$, $N(1535)(1/2^-)$ resonances \cite{gloz96}. We
found also that this structure holds for all $S-$wave baryons
(ground and radially excited baryon states). However the important
question, does this structure hold for orbitally excited baryon
resonances, is still open. Below we show that in the latter case
the one-meson exchange interaction has more complicated structure.
More precisely, it contains additionally an operator which couples the upper
orbital momentum of valence quark emitting a single meson with the
lower orbital momentum of another valence quark absorbing this
meson. In the S-wave limit this operator is proportional to the
spin operator, in other words it supports the above well-known
structure. However, for P-, D- wave baryons the exchange operator
yields different matrix elements and the above structure of the
one-meson exchange does not hold. It is clear that this
relativistic effect can not be obtained in non-relativistic meson
exchange models. As a result we have a splitting due-to one-meson
exchange in the excited baryon sector. This finding concerns the long-standing
"three-body spin-orbit puzzle" in baryons. It is important to note that the obtained
structure of the one-meson exchange operator confirms the prediction of the
large $N_c$ approach. It couples upper and lower orbital momentum, spin,
isospin and full momentum of the two interacting valence quarks.
\par The relativistic quark model is based on an effective chiral Lagrangian describing
quarks as relativistic fermions moving in a confining static potential. The
potential is described by a Lorentz scalar and  the time component of a vector
potential, where the latter term is responsible for short-range fluctuations of
the gluon field configurations \cite{lus81}. The model potential defines
unperturbed wave functions of the quarks which are subsequently
used in the calculations of baryon properties. Interaction of
quarks with Goldstone bosons is introduced on the basis of the
nonlinear $\sigma$-model \cite{gell60}. All calculations are
performed at one loop or at order of accuracy $o(1/f_{\pi}^2)$.
Due to negligible contribution of the  $K-$ and $\eta-$ meson loop
diagrams in our model to the Nucleon and Delta sectors we restrict to the
$\pi-$ meson loop diagrams.
\par In the following we proceed as follows: we
first describe the basic formalism of our approach. Then we
indicate the main derivations relevant to the problem and finally
present the numerical results.

\section{Model}
\par The effective Lagrangian of our model
${\cal L}(x)$ contains the quark core part ${\cal L}_Q(x)$ the quark-pion
 ${\cal L}_I^{(q\pi)}(x)$ interaction part, and the kinetic part for the pion field ${\cal L}_{\pi}(x)$:
\\ \\
\begin{eqnarray}
 \nonumber {\cal L}(x) = {\cal L}_Q(x) + {\cal L}_I^{(q\pi)}(x) +
{\cal L}_{\pi}(x) \\
 = \bar\psi(x)[i\not\!\partial -S(r)-\gamma^0V(r)]\psi(x) - 1/f_{\pi}
 \bar\psi[S(r) i \gamma^5 \tau^i \phi_i]\psi+
   \frac{1}{2}(\!\partial_{\mu}\phi_i)^2-\frac{1}{2}m_i^2\phi_i^2.
\end{eqnarray}
Here, $\psi(x)$ and $\phi_i, i=1,2,3$  are the quark and pion
field operators, respectively. The matrices $\tau^i (i=1,2,3)$
 are the isospin matrices. The pion decay constant is $f_\pi=$93 MeV.
 The scalar part of the static confinement potential
\begin{equation}
S(r)=cr+m
\end{equation}
where c and m are constants. The constant part of the scalar
potential can be interpreted as the current quark mass term.
\par At short distances, transverse fluctuations of the string are dominating
\cite{lus81}, with some indication that they transform like the time component of the
Lorentz vector. They are given by a Coulomb type vector potential as
\begin{equation}
V(r)=-\alpha/r
\end{equation}
where $\alpha$ is approximated by a constant.
The quark fields are obtained from solving the Dirac equation with the corresponding
scalar plus vector potentials
\begin{equation}
[i\gamma^{\mu}\partial_{\mu} -S(r)-\gamma^0V(r)]\psi(x)=0
\end{equation}
The respective positive and negative energy eigenstates as solutions to the Dirac
equation with a spherically symmetric mean field, are given in a general form as
\begin{eqnarray} \label{Gaussian_Ansatz}
 u_{\alpha}(x) \, = \,
\left(
\begin{array}{c}
g^+_{N\kappa }(r) \\
-i f^+_{N\kappa }(r) \,\vec{\sigma}\hat{\vec x} \\
\end{array}
\right)
\, {\cal Y}_{\kappa}^{m_j}(\hat{\vec x}) \,\chi_{m_t} \, \chi_{m_c} \, exp(-iE_{\alpha}t)
\end{eqnarray}

\begin{eqnarray}
 v_{\beta}(x) \, = \,
\left(
\begin{array}{c}
g^-_{N\kappa}(r) \\
-i f^-_{N\kappa}(r) \,\vec{\sigma}\hat{\vec x} \\
\end{array}
\right)
\, {\cal Y}_{\kappa}^{m_j}(\hat{\vec x}) \,\chi_{m_t} \, \chi_{m_c} \, exp(+iE_{\beta}t)
\end{eqnarray}
The quark and antiquark eigenstates $u$ and $v$ are labeled by the radial, angular,
azimuthal, isospin and color quantum numbers $N,\, \kappa,\, m_j,\, m_t$ and $m_c$, which are
collectively denoted by $\alpha$ and $\beta$, respectively. The spin-angular part of the quark
field operators
\begin{equation}
{\cal Y}_{\kappa}^{m_j}(\hat{\vec x})\,=\,[Y_l(\hat{\vec x})\otimes
\chi_{1/2}]_{jm_j} \, \, j=|\kappa|-1/2.
\end{equation}
For a given total angular momentum $j$ and projection $m_j$, the upper and lower
components of Eq.(6) and Eq.(7) are expanded in a harmonic oscillator basis.
The quark fields $\psi$ are expanded over the basis of positive
and negative energy eigenstates as
\begin{equation}
\psi(x)=\sum \limits_{\alpha} u_{\alpha}(x)b_{\alpha} +\sum \limits_{\beta} v_{\beta}(x)d^{\dag}_{\beta} .
\end{equation}
The expansion coefficients $b_{\alpha}$ and $d^{\dag}_{\beta}$ are operators, which
annihilate
a quark and create an antiquark in the orbits $\alpha$ and $\beta$, respectively.
\par The free pion field operator is expanded over plane wave solutions as
\begin{equation}
\phi_j(x)=(2\pi)^{-3/2}\, \int\frac{d^3k}{(2\omega_k)^{1/2}}[a_{j{\bf k}}exp(-ikx)+a^{\dag}_{j{\bf k}}exp(ikx)]
\end{equation}
with the usual destruction and creation operators $a_{j{\bf k}}$
and $a^{\dag}_{j{\bf k}}$ respectively. The pion energy is defined
as \\ $\omega_k \,=\, \sqrt{k^2+m_{\pi}^2}. $
\par In denoting the three-quark vacuum state by $ |0> $, the corresponding
noninteracting many-body quark Green's function (propagator) is given by
the customary vacuum Feynman propagator for a binding potential \cite{fet71}:
\begin{equation}
iG(x,x')\,=\, iG^F(x,x')\,=\,<0|T\{\psi(x) \bar\psi(x')\}|0>\,=\,
\sum \limits_{\alpha} u_{\alpha}(x)\bar u_{\alpha}(x')\theta(t-t') +
\sum \limits_{\beta} v_{\beta}(x)\bar v_{\beta}(x')\theta(t'-t)
\end{equation}
Since the three-quark vacuum state $|0>$ does not contain any
pions , the pion Green's functions are given by the usual free
Feynman propagator for a boson field:
\begin{equation}
i\Delta_{ij}(x-x')\,=\, <0|T\{\phi_i(x) \bar\phi_j(x')\}|0>\,=\,
-\delta_{ij}\int\frac{d^4k}{(2\pi)^4}\frac{1}{k^2-m_{\pi}^2+i\epsilon}
\,exp[-ik(x-x')] \, .
\end{equation}

\par Using the effective Lagrangian we calculate the lowest excitation spectrum
of the nucleon and delta. In the model the quark core result
($E_Q$) is obtained by solving the Dirac equation for the single
quark system numerically. Since we work in the independent
particle model, the bare three-quark state of the $SU(2)$-flavor
baryons has the structure $(1S_{1/2})^2(nlj)$ in the
non-relativistic spectroscopic notation. The corresponding quark
core energy is evaluated as the sum of single quark energies with:
$$ E_Q=2E(1S_{1/2}) + E(nlj)$$
\par The result for $E_Q$ still contains the contribution of the center of mass motion.
To remove this additional term we resort to three approximate
methods, which correct for the center of mass motion: the $R=0$
\cite{lu98}, $P=0$ \cite{teg82} and LHO \cite{wil89} methods. All
these methods were examined in \cite{dong00} for the center of
mass correction of the ground state nucleon and delta and they
give similar results. The first method is based on the extraction
of the center of mass motion using the expression of the baryon
wave function in terms of the Jacobi coordinates and putting $R=0$
in intrinsic wave function (center of mass system). The second
method uses the Fourier transformation of the c.m. coordinates and
setting $P=0$ under Fourier integral. The last method is based on
the keeping the lowest s-state for the center of mass motion in
the product wave function, similar to the non-relativistic shell
model. In present work, we do CM correction using above three methods
for the quark core results of the nucleon and delta ground states. For the
excited states we calculate the splitting from ground states both due-to
quark core and one-pion perturbative corrections.
\par The second order perturbative corrections to the energy spectrum of the
 SU(2) baryons due to pions ($\Delta E^{(\pi)}$)
 are calculated on the basis of the Gell-Mann and Low theorem :
 \begin{eqnarray}\label{Energy_shift}
\hspace*{-.8cm}
\Delta E=<\phi_0| \, \sum\limits_{i=1}^{\infty} \frac{(-i)^n}{n!} \,
\int \, i\delta(t_1) \, d^4x_1 \ldots d^4x_n \,
T[{\cal H}_I(x_1) \ldots {\cal H}_I(x_n)] \, |\phi_0>_{c}
\end{eqnarray}
with $n=2$, where the relevant quark-pion interaction Hamiltonian
density is
\begin{eqnarray}
{\cal H}_I^{(q\pi)}(x)= \frac{i}{f_{\pi}}\bar\psi(x)\gamma^5
\vec\tau\vec\phi(x)S(r)\psi(x),
\end{eqnarray}
The index (c) in Eq.(13) denotes the contributions only from
connected graphs. The stationary bare three-quark state (3-quark core)
 $|\phi_0>$ is constructed from the vacuum state using the usual creation
operators:
\begin{equation}
|\phi_0>_{\alpha\beta\gamma}=b_{\alpha}^+b_{\beta}^+b_{\gamma}^+|0>,
\end{equation}
where $\alpha, \beta$ and $ \gamma$ represent the quantum numbers
of the single quark states, which are coupled to the respective
baryon configuration. The energy shift of Eq.(13) is evaluated up
to second order in the quark-pion interaction, and generates
self-energy and exchange contributions.

\subsection{ Self-energy contributions}
\par The self-energy terms contain contributions both from
intermediate quark
$(E>0)$ and antiquark $(E<0)$ states.
\par The self energy term due-to pion fields (see Fig.1) is evaluated as
\begin{eqnarray}
\Delta
E_{s.e.}^{(\pi)}=-\frac{1}{2f_{\pi}^2}\sum\limits_{a=1}^{3}\sum\limits_{\alpha
  ' \leq \alpha_F} \int
\frac{d^3\vec p}{(2\pi)^3p_0} \biggl\{ \sum\limits_{\alpha}\frac{V_{\alpha
      \alpha ' }^{a+}(\vec p)V_{\alpha \alpha ' }^{a}(\vec
        p)}{E_{\alpha}-E_{\alpha '}+p_0}-
 \sum\limits_{\beta}\frac{V_{\beta \alpha '}^{a+}(\vec p) V_{\beta
 \alpha '}^{a}(\vec p)}{E_{\beta}+E_{\alpha '}+p_0}\biggr\},
\end{eqnarray}
with $ p_0^2=\vec p^2 + m_{\pi}^2$.
The transition form factors are defined by:
\begin{eqnarray}
V_{\alpha\alpha'}^a(\vec p)=\int d^3 x \bar u_{\alpha}(\vec x)\Gamma^a(\vec x)
u_{\alpha '}(\vec x) e^{-i\vec p \vec x} \\
V_{\beta \alpha'}^a(\vec p)=\int d^3 x \bar v_{\beta}(\vec x)\Gamma^a(\vec x)
u_{\alpha '}(\vec x) e^{-i\vec p \vec x}
\end{eqnarray}
The vertex function of the $\pi -q -q $ and $\pi -q -\bar q $
transition is
\begin{eqnarray}
\Gamma^a= S(r) \gamma^5 \tau^a I_c \, ,
\end{eqnarray}
where $I_c$ is the color unity matrix. The sum in Eq.(16) is performed over
$\alpha '$ up to and including the Fermi level
with quantum number $\alpha_F$ and over all quantum
numbers $\alpha$ and $\beta$ of the intermediate quark state with
it's positive and negative energy solutions. After estimation of
the transition form factors (see Appendix) and putting into
equation (16) and integration over angular part in the momentum
space, we obtain next expression for the energy shift of the SU(2)
baryon state due to the second order self-energy diagrams:
\begin{eqnarray}
\nonumber \Delta E_{s.e.}^{(\pi)}=-\frac{1}{16\pi^3f_{\pi}^2}\int
\frac{d p \, p^2}{p_0} \sum\limits_{N',l',j'}\sum\limits_{l_n}
\biggl\{ \sum\limits_{\alpha}\frac{[\int dr r^2 G_{\alpha \alpha
    '}(r)S(r)j_{l_n}(pr)]^2}{E_{\alpha}-E_{\alpha '}+p_0}
Q_{s.e.}(l,l',l_n,j,j') - \\
\sum\limits_{\beta}\frac{[\int dr r^2
G_{\beta \alpha '}(r)S(r)j_{l_n}(pr)]^2}{E_{\beta}+E_{\alpha '}+p_0}
Q_{s.e.}(l,l',l_n,j,j') \biggr \} ,
\end{eqnarray}
where $j_{l_n}$ is the Bessel function. The radial overlap of the
single quark states with quantum numbers
$\alpha=(N,l,j,m_j,m_t,m_c)$ and $\alpha^{\prime}=(N',l',j',m_j
',m_t ',m_c ')$ is defined as
\begin{eqnarray}
  G_{\alpha \alpha '}(r)=f_{\alpha}(r)g_{\alpha '}(r) + f_{\alpha '}(r)
g_{\alpha }(r)  .
\end{eqnarray}
The angular momentum coefficients $Q$ are
evaluated for all SU(2) baryons as
\begin{eqnarray}
Q_{s.e.}(l,l',l_n,j,j')= 12\pi [l^{\pm}][l_n][j] \biggl [
C^{l'0}_{l^{\pm}0l_n 0} W (j \frac{1}{2}l_nl'; l^{\pm}j')\biggr ]^2
\sum\limits_{m_j}\sum \limits_{m_j ' \leq \alpha_f} \biggl [
C^{j'm_j'}_{jm_jl_n(m_j'-m_j)} \biggr ]^2 ,
\end{eqnarray}
where $C$ and $W$ are the Clebsch-Gordan and Wigner coefficients,
respectively. The sum in Eq.(22) does not depend on the
orientation of the full momentum of the valence quark $m_j
'=-j',-j'+1,...,j'-1,j'$. In the case of the ground state N and
$\Delta$ the sum over parameter-set $(N',l',j')$ is replaced by
the factor 3, since a valence quark in this case can be only in
$1S_{1/2}$ state. The first term in the sum of Eq.(20) represents
the contribution from intermediate quark states, and the second
term corresponds to the contribution of intermediate antiquark
states to the energy shift of the SU(2) baryon state.
\par We note also that the role of meson cloud (self-energy) corrections
can be also pinned down by sigma-terms and when considering the chiral limit
\cite{inoue041}.
\begin{figure}[tbh]
\begin{center}
\includegraphics[width=15cm]{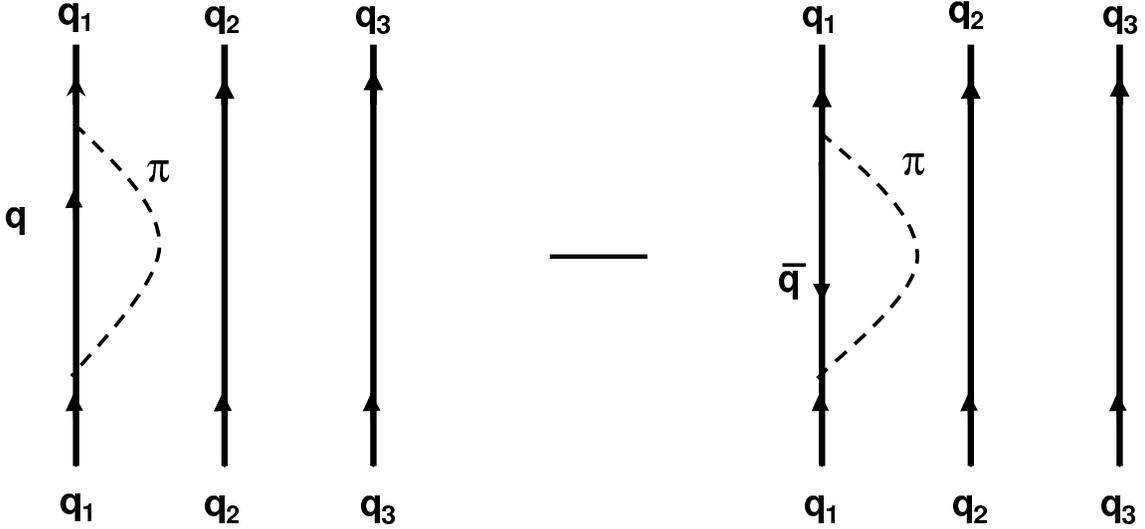}
\end{center}
\caption{Second order self energy diagrams due-to $\pi-$meson
fields \label{Fig1}}
\end{figure}
\subsection{ Exchange diagrams contribution}
 \par The exchange term due-to pion fields (see Fig.2) is evaluated as:
 \begin{eqnarray}
\Delta
E_{ex.}^{(\pi)}=-\frac{1}{2f_{\pi}^2}\sum\limits_{a=1}^{3}\sum\limits_{\alpha
 \leq  \alpha_F}\sum\limits_{\alpha ' \leq \alpha_F} \int\frac{d^3\vec p}
{(2\pi)^3p_0^2} \biggl\{ V_{\alpha \alpha  }^{a+}(\vec p)
V_{\alpha '\alpha ' }^{a}(\vec p)- V_{\alpha \alpha '}^{a+}(\vec p)
V_{\alpha  \alpha '}^{a}(\vec p) \biggr\}.
\end{eqnarray}
 By using the explicit form of the transition form factor (see Appendix)
 and Wick's theorem we can write  more convenient  expression for the
 energy shift due to the second order pion exchange
 diagrams:
  \begin{eqnarray}
\Delta E_{ex.}^{(\pi)}=-\frac{1}{16\pi^3f_{\pi}^2}\int \frac{d p
\, p^2}{p_0^2} \sum\limits_{l_n}\hat{\Pi}_{l_n}(p)
\end{eqnarray}
where
\begin{eqnarray}
\hat{\Pi}_{l_n}(p)=<\phi_B|\sum\limits_{i\neq j}
\hat{\vec\tau}(i)\hat{\vec\tau}(j)\hat{T}_{ln}(i)
\hat{T}_{l_n}(j)\hat{K}_{l_n}(i)\hat{K}_{l_n}^+(j)|\phi_B> \,
\end{eqnarray}
and the operators $\hat{\vec\tau}$, $\hat{T}_{l_n}$ and $
\hat{K}_{l_n}$ are summed over single quark levels $i\neq j$ of
the SU(2) baryon. In the quark model, the baryon wave function $|\phi_B> $
is presented as a bound state of three quarks, and it can be written down commonly
as
\begin{eqnarray}
\nonumber
\|\phi_B>=|\alpha\beta\gamma>=\sum
\limits_{J_0T_0}|\alpha\beta;\gamma>_{JM(J_0)}^{TM_T(T_0)} \\
=\sum\limits_{J_0T_0}\hat {S}\biggl [ |\psi_{\alpha}(r_1)\psi_{\beta}(r_2)
\psi_{\gamma}(r_3){\cal Y}_{J_0}^{JM}(\hat{x_1}\hat{x_2};\hat{x_3})>
|\chi_{T_0}^{TM_T}(12;3)>|\chi_c^(123)>\biggr ],
\end{eqnarray}
where $J_0$ and $T_0$ are intermediate spin and isospin couplings
respectively. The states $\psi$ are the single particle states,
labeled by a set of quantum numbers $\alpha$, $\beta$ and $\gamma$,
excluding the color degree of freedom.
\par The operator  $\hat{T}_{l_n}$ in Eq.(25) is the radial integration
operator with the factor $j_{l_n}(pr)S(r)$:
\begin{eqnarray}
<\alpha|\hat{T}_{l_n}|\beta>=\int dr\biggl [r^2
S(r)j_{l_n}(pr)G_{\alpha\beta}(r)\biggr ].
\end{eqnarray}
The matrix elements of the operator $\hat{K}_{l_n}$ are given by
\begin{eqnarray}
\nonumber
<\alpha|\hat{K}_{l_n}|\beta>=-\biggl( 4\pi
[l^{\pm}(\alpha)][l_n][j(\alpha)]\biggr)^{1/2}C^{l(\beta)0}_{l^{\pm}
(\alpha)0l_n 0}
  W (j(\alpha)\frac{1}{2}l_n,l(\beta); l^{\pm(\alpha)},j(\beta))
  \\
C^{j(\beta)m(\beta)}_{j(\alpha)m(\alpha)l_n(m(\beta)-m(\alpha))},
\end{eqnarray}
where $j(\alpha), l(\alpha), l^{\pm}(\alpha), m(\alpha)$ and
$j(\beta), l(\beta), l^{\pm}(\beta), m(\beta)$ are the quantum
numbers of the single quark states $<\alpha|$  and  $<\beta|$: the
full momentum, upper orbital momentum, lower orbital momentum, and
projection of the full momentum respectively. The Hermitian
conjunction operator is defined as
\begin{equation}
<\alpha|\hat{K}_{l_n}^+|\beta>=<\beta|\hat{K}_{l_n}|\alpha>.
\end{equation}
It is important to note that operator $\hat{K}_{l_n}$ couples the
upper component of the single quark state $\alpha$ (which
corresponds to the valence quark emitting a single pion) with the
lower component of the single quark state $\beta$ (which
corresponds to the valence quark absorbing this pion). And
operator $\hat{K}_{l_n}^+$ acts vice versa. These operators
correspond to the coefficient ${\cal F}$ in Appendix. It is clear
that they define the relativistic structure of the one-meson
exchange. For the S-wave baryons, i.e. when $\alpha$ and $\beta$
single quark states are S-quarks, the operators $\hat{K}_{l_n}(i)$
and $\hat{K}_{l_n}^+(j)$ are proportional to the spin operators
$\hat{\sigma}_i$ and $\hat{\sigma}_j$ respectively. It means that
for the S-wave baryons the relativistic and non-relativistic
structure of one-meson exchange are the same. However for the
orbitally excited baryon states (P-, D- wave baryons) they are
different due-to operator $\hat{K}_{l_n}$ which couples the lower
and upper components of the interacting quarks.
\begin{figure}[tbh]
\begin{center}
\includegraphics[width=15cm]{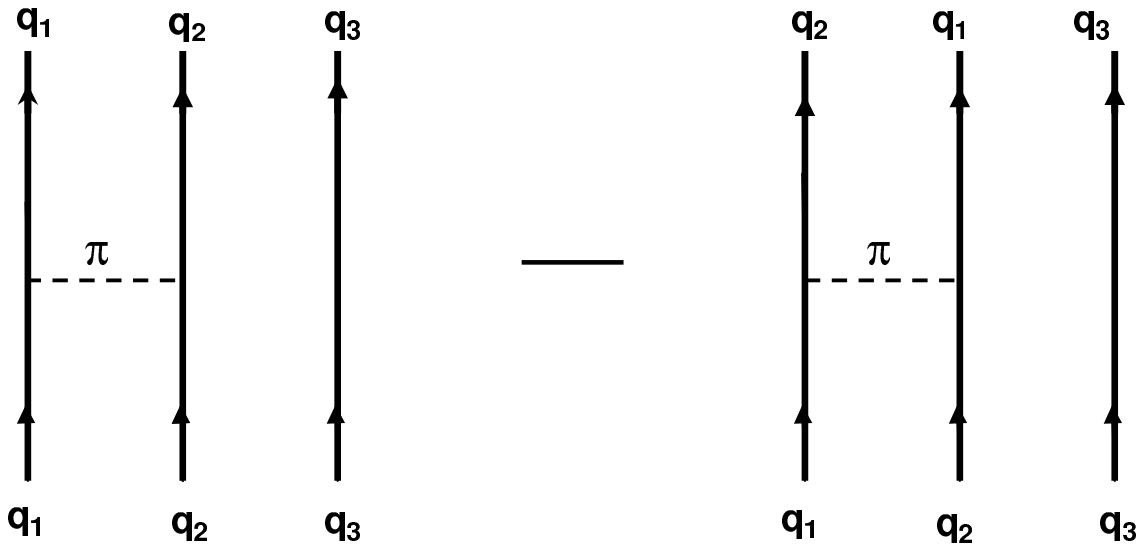}
\end{center}
\caption{Second order exchange diagrams due-to $\pi-$meson
fields \label{Fig2}}
\end{figure}

\section{Numerical results}

\par In order to account for the finite size effect of the pion we introduce
a one-pion vertex regularization function in the momentum space,
parameterized in the dipole form as
$$ F_{\pi}(p^2)= \frac {\Lambda_{\pi}^2-m_{\pi}^2}{\Lambda_{\pi}^2+p^2}.$$
According to the study of long-wavelength transverse fluctuations
of the flux-tube \cite{lus81} we use the value $\alpha=0.26\approx
\pi/12$. We choose two sets of parameters: model A and model B.
 The parameters of the confining potential ($c$ and $m$) are fitted to
 reproduce correct axial charge of the proton which remains unaltered when
 including the effects of the pion cloud, the
empirical value of the pion-nucleon coupling constant
$$G^2_{\pi N N} /4\pi \approx 14$$ and a normal value for the quark core
RMS radius $0.5$ fm of the proton. Thus, we use only few parameters to reproduce
the SU(2) baryons spectrum contrary to the traditional OGE and GBE based quark models.
\par In Table 1 we indicate the model parameters together with the corresponding
single-quark energies. We note here that the first orbitally
excited single valence quark level is $1P_{3/2}$. Therefore we
associate the first orbitally excited nucleon and delta states
with the structure $(1S)^21P_{3/2}$.
 \par Table 2 contains the quark core results \cite{gutsche87} for the static properties
 of the proton. The results include the center of mass correction.
 After CM correction, as was shown in \cite{gutsche87}, the magnetic moment of the proton
 increases by about  20$\%$ , and the axial charge decreases by about 5$\%$.
 The CM correction reduces the RMS radius of the proton approximately by 10$\%$ of the
 uncorrected value. From the table one can see, that a larger value of the strength parameter
  $c$ of the confining potential yields a smaller value for the RMS radius of the proton. An
 estimation of the pion-nucleon coupling constant is close to the
 empirical value for both Model A and Model B.
 \par Perturbative corrections to the lower SU(2) baryon states energy
shift values due-to self-energy diagrams are given in Table 3.
Contribution from intermediate quark and antiquark states are
taken with the full momentum up to $j=25/2$. As was shown in
\cite{gutsche87} for the ground state nucleon, the pion
self-energy contribution is positive and convergent also for
excited Nucleon and Delta states in contrast to the bag models.
Contributions from self-energy diagrams for the radially excited
Nucleon and Delta states with the last valence quark in $2S_{1/2}$
state are larger than for excited Nucleon and Delta states with
the last valence quark in orbitally excited $1P_{3/2}$  and
$1P_{1/2}$ states.
\par Corresponding energy shift values due-to exchange diagrams are given
in Table 4. As we noted above, the relativistic structure of
one-meson exchange operator results a splitting between negative
parity nucleon resonances and also between delta resonances. The
nucleon states with the structure $(1S)^21P_{3/2}$ split according
to the energy shift values -132.8 and -150 MeV. In contrast to the
GBE based quark models, lowest radial excitation of the Nucleon
$N^*(1440)(1/2^+)$ which has one valence quark in $2S_{1/2}$ state,
in our model is less bound with 120.35 MeV  than the lowest
orbital excitations of the Nucleon $N^*(1520)(3/2^-)$ and
$N^*(1535)(1/2^-)$, with 132.8 and 150 MeV respectively, which
have the last valence quark in $1P_{3/2}$ state.
\par In Table 5 we give the mass values for the g.s. N(939) with and without CM
correction for both parameter sets A and B. In the case of the
model A, a reasonable  value (1280-1320 MeV) for the mass of the
$\Delta$ is obtained. However, for the nucleon ground state the
estimation is still large. Model B yields too large value for the
mass of the nucleon. This means that larger values of the strength
parameter $c$ of the confining potential is not likely.
From the Table one can see that the three methods for the correction of center
of mass motion agree within 50 MeV which seems too large. However, these
three methods always give corrections with systematic differences. For example,
the LHO method yields correction larger than the $P=0$ method, but smaller than
the $R=0$ method. Thus, we can fix one of these methods and analyse the excited
sector.
\par And finally in Table 6 we compare
the theoretical energy splitting values with experimental data
from PDG \cite{PDG03} and recent analysis \cite{arndt04}. The new
analysis of Arndt et al. does not contain the resonance
$\Delta(1600)(3/2^+)$. We note that the results presented in the last table do
not include any correction due-to CM motion. The quark core and self-energy contributions
to the Nucleon and Delta states with the same structure (for example $(1S)^2(1P_{3/2})$)
are equal. Therefore the contributions from quark core and self-energy terms to the
$N(939) - \Delta(1232)$, $N(1520) - N(1535)$ and $\Delta(1620) - \Delta(1700)$ splitting
are exactly zero. The only correction due-to CM motion in the excited Nucleon sector in
Table 6 is needed for $N(939) - N(1440)$ and $N(1440) - N(1535)$ splitting. However,
we can estimate this correction by using an approximate factor
from $1/2$ to $2/3$. First of all we note a good agreement of the
theoretical value for the splitting of the first
 radial excitation of the Nucleon, Roper resonance
 $N^*(1440)(1/2^+)$. The quark core result 415 MeV in the Model A should restrict
 to about 200-250 MeV after CM correction. That gives a good
 agreement with the experimental data when adding a perturbative correction
 (284 MeV for the model A). However the ordering and splitting of the
 Roper resonance and orbital Nucleon excitations $N^*(1520)(3/2^-)$ and
$N^*(1535)(1/2^-)$ is not reproduced in our model contrary to the
non-relativistic Constituent Quark Model of Glozman et al.
\cite{gloz96}. We suggest that the difference
comes mainly from the relativistic structure of the one-meson exchange.
\par We also note that the splitting between Delta states is well
reproduced though the last analysis of Arndt et al. shows some
different picture.

\section{Summary and conclusions}
\par We demonstrated that a relativistic chiral quark model can
yield a reasonable description of the energy spectrum of the
ground state $N$ and $\Delta$ and their lowest radial and orbital
excitations. The model yields a good splitting of the first
radial excitation of the nucleon, the Roper resonance $N^*(1440)$
from ground state $N(939)$ and splitting between negative parity
nucleon $N^*(1520)(3/2^-)$ and  $N^*(1535)(1/2^-)$ resonances. The
splitting in the $\Delta$ sector also was reproduced. It is
important to note that these results correspond to the parameters
of the model, which yield a normal quark core radius of the
nucleon, being about 0.5 fm and empirical value of the
pion-nucleon coupling constant $G^2_{\pi N N} /4\pi \approx 14$.
\par We found that the relativistic structure of one-meson exchange
interaction confirms a prediction of the large $N_c$ approach \cite{schat02},
which stated that the mass splitting between various baryons receive important
contributions from operators which simultaneously couple spin, isospin and
orbital momentum. In particular, the one-pion exchange interaction
includes an operator which couples the lower and
upper orbital momentum of two interacting valence quarks. In the
S-wave limit (ground and radially excited Nucleon and Delta
states) this operator is proportional to the simple  spin
operator. However, for the orbitally excited baryon resonances it
yields different matrix elements and splitting between for example
negative parity nucleon resonances and delta resonances.
 As a result we conclude that relativistic one-meson exchange
interaction does not yield  the correct ordering of lowest
positive and negative parity nucleon resonances and a new mechanism is needed
for the explanation of this effect.
\par Since the developed model gives only a part of the $N(939)-
\Delta(1232)$ mass splitting and that the Delta states are
reproduced quite well, we see two possible development of the
model. The first is to include Instanton induced exchange
mechanism, which should give another part of the $N(939)-
\Delta(1232)$ splitting. It seems that these forces are
responsible for the correct ordering of the radially and orbitally
excited Nucleon resonances. On the other hand, they don't change
the spectrum of the Delta states and thus keep a good
description of this sector. The second development would include
the one-gluon exchange forces which should give some part of the
above splitting, and possible inclusion of the Instanton induced
interaction mechanisms. The obtained splitting between negative
parity nucleon resonances due-to relativistic one-meson exchange
forces in our model would cancel a large value of the spin-orbit
interaction due-to one-gluon exchange. This would help to
understand a long standing "three body spin-orbit puzzle" in
baryons.
\\
{\it Acknowledgements}. This work  was supported in part by the
DAAD research fellowship (Germany) and by the Belgian-state
Federal Services for Scientific, Technical and Cultural Affairs
(SSTC). The author acknowledges the Institut
f\"ur Theoretische Physik, Universit\"at T\"ubingen, the PNTPM
group of ULB, Brussels and the Abdus Salam International Center
for Theoretical Physics for kind hospitality. He deeply thanks
Prof. A. Faessler, Prof. D. Baye and Dr. Th. Gutsche for valuable advices
and discussion, Prof. F. Stancu, Dr. V.~E.~Lyubovitskij and Dr. A. Rakhimov for useful
discussions and Prof. B.S. Yuldashev and Prof. U.S. Salikhbaev for
their interest.

\clearpage
\begin{center}
    {\Large Appendix: Transition form factors}
\end{center}
\vskip 0.5cm
\par Putting explicit expression of the vertex matrix
$\Gamma^a(\vec x)$ from Eq.(19) into Eq.(17) we receive next
equation:
\begin{eqnarray}
\nonumber
V_{\alpha\alpha'}^a(\vec p)&=& -i \int dr
r^2\Big[g_{\alpha}(r)f_{\alpha'}(r)+
g_{\alpha'}(r)f_{\alpha}(r)\Big]S(r) \\
&   &   \int d\hat{r}\Big[ {\cal
Y}_{jl}^{m_j^+}(\hat{r})(\vec{\sigma}\hat{r}) {\cal
Y}_{j'l'}^{m_j'}(\hat{r})e^{-i\vec{p}\vec{r}}\Big]
<m_t|\tau^a|m_t'><m_c|I_c|m_c'>
\end{eqnarray}
Now using
$${\cal Y}_{jl}^{m_j^+}(\hat{r})(\vec{\sigma}\hat{r})=-{\cal Y}_{jl^{\pm}}^{m_j^+}(\hat{r})
$$
which couples the lower orbital momentum to the spin, and
expanding the exponential function over spherical Bessel functions
and integrating over angular part of the variable $\vec r$, we get
next equation for the integral
$$\int d\hat{r}\Big[ {\cal
Y}_{jl}^{m_j^+}(\hat{r})(\vec{\sigma}\hat{r}) {\cal
Y}_{j'l'}^{m_j'}(\hat{r})e^{-i\vec{p}\vec{r}}\Big]=
\sum_{l_n}(-i)^{l_n} j_{l_n}(pr) Y_{l_n}^{m_j'-m_j}(\hat{p})
 {\cal F}(l^{\pm},l',l_n,j,j',m_j,m_j'),
 $$
 where coefficients ${\cal F}$ are defined as
 $$ {\cal F}(l^{\pm},l',l_n,j,j',m_j,m_j')=
-\biggl( 4\pi [l^{\pm}][l_n][j]\biggr)^{1/2}C^{l'0}_{l^{\pm}0l_n0}
  W (j\frac{1}{2}l_n,l'; l^{\pm},j')
C^{j'm'_j}_{jm_jl_n(m'_j-m_j)}.
$$
For the transition form-factor now it is easy to write the next
expression:
\begin{eqnarray}
\nonumber
V_{\alpha\alpha'}^a(\vec p) & = & \sum_{l_n} (-i)^{l_n+1}\int dr
r^2\Big[g_{\alpha}(r)f_{\alpha'}(r)+
g_{\alpha'}(r)f_{\alpha}(r)\Big]S(r) j_{l_n}(pr) \\
 &  &  Y_{l_n}^{m_j'-m_j}(\hat{p}) {\cal F}(l^{\pm},l',l_n,j,j',m_j,m_j')
<m_t|\tau^a|m_t'><m_c|I_c|m_c'>.
\end{eqnarray}
The Hermitian conjunction of the transition form factor
\begin{eqnarray}
\nonumber
V_{\alpha\alpha'}^{a+}(\vec p)& = & \sum_{l_n} (i)^{l_n+1}\int dr
r^2\Big[g_{\alpha}(r)f_{\alpha'}(r)+
g_{\alpha'}(r)f_{\alpha}(r)\Big]S(r) j_{l_n}(pr) \\
 &  &  Y_{l_n}^{(m_j'-m_j)*}(\hat{p}) {\cal
F}(l^{\pm},l',l_n,j,j',m_j,m_j') <m_t'|\tau^a|m_t><m_c'|I_c|m_c>.
\end{eqnarray}

\newpage
%
\begin{table}
\caption { Parameter sets for the models A and B and corresponding
single quark energies in MeV}
\begin{tabular}{|c|c|c|c|c|c|c|c|c|} \hline
 Model          & c, Gev$^2$ &  m, Gev & $\Lambda_{\pi}$, Gev & $\alpha$
                & $E(1S)$ & $E(2S)$ &  $E(1P_{3/2})$ & $E(1P_{1/2})$       \\    \hline
   A        & 0.16  & 0.06 &1.0 &0.26 & 571.7&986.7& 822.8 & 860.7          \\  \hline
   B        & 0.20  & 0.07 & 1.2 & 0.26 & 641.4 & 1105.71 &922.0 & 964.4     \\  \hline
\end{tabular}
\caption {Quark core contributions to the static properties of
the proton \cite{gutsche87} }
\begin{tabular}{|c|c|c|c|c|}  \hline
 Model & $g_A$  &$\mu_p, N.M.$  & RMS radius, fm & $G^2_{\pi NN}/(4\pi )$   \\    \hline
 A     & 1.26   & 1.58   &  0.52   & 13.919                    \\      \hline
 B     & 1.26   & 1.41   &  0.47   & 13.984                    \\  \hline
 \end{tabular}
\caption { Second order perturbative corrections due to one-pion
 self energy diagrams for the energy shift of the single valence quarks in MeV }
 \begin{tabular}{|c|c|c|c|c|}  \hline
   Model     & 1S           & 2S  &$1P_{3/2}$ & $1P_{1/2}$  \\   \hline
   A         & 126.5        & 350 &  262      & 248       \\   \hline
   B         & 193          & 540 &  400      & 380       \\  \hline
\end{tabular}
\caption { Second order perturbative corrections due to one-pion
 exchange diagrams for the mass spectrum of lowest
 $N$ and $\Delta$ states in MeV for the Model A}
 \begin{tabular}{|c|c|c|c|c|c|c|}  \hline
 (J,T)&$(\frac{1}{2},\frac{1}{2})$& $(\frac{3}{2},\frac{1}{2})$
 &$(\frac{5}{2},\frac{1}{2})$&
 $(\frac{1}{2},\frac{3}{2})$ & $(\frac{3}{2},\frac{3}{2})$&
 $(\frac{5}{2},\frac{3}{2})$   \\ \hline
 $(1S)^3$         & -179.5  &     &  &     & -35.9 &   \\   \hline
 $(1S)^22S$       & -120.35 &     &  &     & -24   &   \\  \hline
 $(1S)^21P_{1/2}$ & 3.7     &-69.1&  &-23.8&-17.4  & \\  \hline
 $(1S)^21P_{3/2}$ &-132.8&-150 & 10.8 & -7.5 &-17.6 & -34.5 \\  \hline
\end{tabular}
\caption { The mass value of the g.s. nucleon in MeV  with and
without center of mass (CM) correction}
 \begin{tabular}{|c|c|c|c|c|c|}  \hline
 Model &              & No CM   & R=0, \cite{lu98}& P=0, \cite{teg82}& LHO, \cite{wil89}   \\    \hline
 A     & $E_Q$        & 1715    & 940             & 985              & 966        \\
 & $E_Q+\Delta E$     & 1915    & 1140            & 1185             & 1166        \\  \hline
 B & $E_Q$            & 1924    & 1057            & 1110             & 1088         \\  \hline
  & $E_Q+\Delta E$    & 2225    & 1358            & 1411             & 1389       \\   \hline
\end{tabular}
\caption { Energy splitting values between lowest $N$ and $\Delta$
states in MeV }
 \begin{tabular}{|c|c|c|c|c|c|c|}  \hline
   & A, $E_Q$ & A, $\Delta E$ & B, $E_Q$ & B, $\Delta E$
   & exp.\cite{PDG03} & exp.\cite{arndt04} \\  \hline
$N(939)(1/2^+)-\Delta(1232)(3/2^+)$ & 0 & 144 & 0 & 223 & 293 &293   \\
\hline $N(939)(1/2^+)-N(1440)(1/2^+)$ & 415 & 284 & 464 & 439
  &$490 \div 530$ & $528 \pm 4.5 $ \\  \hline
 $N(1440)(1/2^+)-N(1535)(1/2^-)$ & -164 &
-101 & -184 & -161 & $50 \div 125 $& $78 \pm 6.5$   \\  \hline
$N(1520)(3/2^-)-N(1535)(1/2^-)$ & 0 & 17 & 0 & 26 &$ 0 \div 30$ &
$30.4 \pm 3$ \\ \hline
 $\Delta(1232)(3/2^+)-\Delta(1600)(3/2^+)$ & 415 & 237 &
464 & 366 & $320 \div 470$ &  \\  \hline
$\Delta(1232)(3/2^+)-\Delta(1620)(1/2^-)$ & 251 & 164 & 280 & 250
&$380 \div 445$ & $381 \pm 2$ \\  \hline
$\Delta(1620)(1/2^-)-\Delta(1700)(3/2^-)$ & 0 &-10 & 0 &-15 &
$(-5) \div 185$ & $74\pm 3.5 $ \\  \hline
\end{tabular}
\end{table}

\begin{thebibliography}{99}
\bibitem{nac03} T. Nacano, et al. [LEPS collaboration], Phys. Rev. Lett. {\bf
91},012002 (2003)
\bibitem{alt04} C. Alt, et al. [NA49 collaboration], Phys. Rev. Lett. {\bf
92}, 042003(2004)
\bibitem{jaf04} R. Jaffe and F. Wilczek, Europ. Phys. Jour. {\bf
C33}, S38-S42(2004)
\bibitem{li91} Z.P. Li, Phys. Rev. {\bf D44}, (1991) 2841
\bibitem{isgur77} N. Isgur and G. Karl, Phys. Lett. {\bf 72B}, 109(1977).
\bibitem{cap86} S. Capstick and N. Isgur, Phys. Rev. {\bf D34},
2809(1986)
\bibitem{stan90} F. Stancu and P. Stassart, Phys. Rev. {\bf D41},
916(1990)
\bibitem{stan91} F. Stancu and P. Stassart, Phys. Lett. {\bf B269},
243(1991)
\bibitem{gloz96} L.Y. Glozman and D.O. Riska, Phys. Rept. {\bf268},
263(1996); L.Y. Glozman, W. Plessas, K. Varga and R.F. Wagenbrunn,
Phys. Rev. {\bf D58} (1998) 094030
\bibitem{isgur99} N. Isgur, Phys. Rev. {\bf D60}, 054013(1999)
\bibitem{cap00} S. Capstick and W. Roberts, nucl-th/0008028, 2000.
\bibitem{PDG03} K. Hagiwara et al. (Particle Data Group), Phys.Rev.
{\bf D66}, 010001(2002)
\bibitem{arndt04} R.A. Arndt, W.J. Briscoe, I.I. Strakovsky, R.L. Workman and M.M. Pavan
Phys. Rev. {\bf C69}, 035213(2004)
\bibitem{MIT} A. Chodos, R. Jaffe, K. Johnson, C.B. Thorn and V. Weisskopf
 Phys. Rev. {\bf D9}, 3471(1974)
\bibitem{thomas81} A.W. Thomas, S. Theberge and G.A. Miller,
Phys. Rev. {\bf D22}, 2838(1980); {\bf D24}, 216(1981)
\bibitem{bro80} G.E. Brown, M. Rho and V. Vento, Phys. Lett. {\bf 97B}, 423(1980)
\bibitem{saito84} K. Saito, Prog. of Theor. Phys. {\bf V71},
775(1984)
\bibitem{dor90} A.E. Dorokhov and N.I. Kochelev, Sov. Jour. Nucl. Phys.{\bf 52},
135(1990)
\bibitem{al99} R.~Alkofer, S.~Ahlig, C.~Fischer, M.~Oettel and H.~Reinhardt;
 Nucl. Phys. A {\bf 663}, 683 (2000); R.~Alkofer, A.~H\"oll, M.~Kloker,
A.~Krassnigg and C.D.~Roberts, Few Body Syst.{\bf 0},1 (2004)
\bibitem{mil02} G.A. Miller, Phys. Rev. {\bf C66}, 032201(2002)
\bibitem{hr05} H.H. Matevosyan, G.A. Miller and A.W. Thomas, nucl-th/0501044
\bibitem{lein04} D.B. Leinweber, W. Melnitchouk, D.G. Richards, A.G. Williams and
J.M. Zanotti, nucl-th/0406032
\bibitem{oset84} E.~Oset, R.~Tegen, and W.~Weise;  Nucl. Phys. A {\bf 426}, 456 (1984)
\bibitem{gutsche87} T.~Gutsche, Ph.~D.~Thesis, Florida State University, 1987
(unpublished).
\bibitem{gutsche89} T.~Gutsche and D.~Robson, Phys. Lett. {\bf B229}, 333 (1989)
\bibitem{lyub01} V.~E.~Lyubovitskij, T.~Gutsche, A.~Faessler, and
E.~G.~Drukarev, Phys. Rev. {\bf D63}, 054026 (2001); V. E.
Lyubovitskij, T. Gutsche and A. Faessler, Phys. Rev. C {\bf 64},
065203 (2001);
\bibitem{inoue041} T. Inoue, V.~E.~Lyubovitskij, T.~Gutsche and A.~Faessler,
 hep-ph/0404051
\bibitem{inoue042} T. Inoue, V.~E.~Lyubovitskij, T.~Gutsche and A.~Faessler,
 hep-ph/0407305
\bibitem{schat02} C.L. Schat, J.L. Goity and N.N. Scoccola, Phys. Rev. Lett.
{\bf 88}, 102002(2002)
\bibitem{lus81} M.~L\"{u}scher, Nucl. Phys. {\bf B180}, 317 (1981).
\bibitem{gell60} M. Gell-Mann and M. Levy, Nuovo Cim. {\bf 16} (1960) 1729
\bibitem{fet71} A.I. Fetter and J.D. Waleska. Quantum theory of many particle systems
(McGraw-Hill, New York, 1971)
\bibitem{lu98} D.H.Lu, A.W.Thomas and A.G.Williams,
Phys.Rev.C57(1998), 2628
\bibitem{teg82} R.Tegen, R.Brockmann and W.Weise, Z.Phys. A307(1982)339
\bibitem{wil89}L.Wilets, "Non-Topological Solitons (World Scientific,
Singapoure, 1989
\bibitem{dong00} Y.B. Dong, K. Shimizu, A. Faessler and A.J. Buchmann,
 Phys. Rev. {\bf C60}, (1999) 035203
\end{thebibliography}
\end{document}